\documentclass[pre,superscriptaddress]{revtex4}
\usepackage{latexsym}
\usepackage{graphicx}
\usepackage{epsfig,subfigure}
\begin{document}
\title{Unified ``micro''- and ``macro-'' evolution of eco-systems:\\
Self-organization of a dynamic network}

\author{Dietrich Stauffer{\footnote{E-mail: stauffer@thp.uni-koeln.de}}}
\affiliation{Institute for Theoretical Physics, Cologne University, D-50923 K\"oln, Euroland.} 

\author{Debashish Chowdhury{\footnote{E-mail: debch@iitk.ac.in}}} 

\affiliation{Department of Physics, Indian Institute of Technology, Kanpur 208016, India.}

\begin{abstract} 
Very recently we have developed a {\it dynamic network} model for 
eco-systems that achieved ``unification'' of ``micro'' and 
``macro''-evolution. We now propose an extension of our model  so 
as to stabilize the eco-system and describe {\it speciation} in a more 
realistic manner. 

\end{abstract}

\maketitle
\bigskip


\section{Introduction}

It is now widely believed that functional networks abound in biological 
systems \cite{hopfield,bornholdt}, from molecular level (e.g., genetic 
and metabolic networks) to the levels of cells, organisms and species. 
An eco-system may be viewed as a functional network of species where 
the (directed) links of the network represent the inter-species 
interactions \cite{foodweb2}. These interactions include not only 
prey-predator interactions but also competitions between species, for the 
same resources, as well as possible cooperations. Each species is also 
a network of individual organisms; the intra-species interactions can 
be cooperative or competitive. The problems of evolutionary ecology 
have been investigated extensively from the perspective of statistical 
physics \cite{drossel,newman,lassig}. 

In recent papers \cite{csk,cspre} we have developed a model from the 
perspective of dynamic networks, by incorporating both inter-species 
and intra-species interactions, for studying some generic features of 
the biological evolution of eco-systems. In this paper, following a 
brief review of this model and its main results, we propose an extension 
of the model so as to capture speciation in a more realistic manner.

\section{Earlier models and their limitations} 

For convenience, most of the earlier theories focussed attention either 
on ``micro''- evolution on ecological time scales (e.g., annual 
variations in the populations of species, mortality rates, etc.) or on 
``macro''-evolution that is most prominent on geological times scales 
(e.g., the phenomena of speciation or extinction). The ecological models 
\cite{murray} that describe population dynamics using, for example, 
Lotka-Volterra type equations \cite{lotka}, usually ignore 
macro-evolutionary changes in the eco-system. On the other hand, most of 
the macro-evolutionary models \cite{kauffman,bak,solerev} do not 
explicitly explore the ageing and age-distributions of the populations 
of various species in the system. The models of ageing and dynamics of 
age-distributed populations \cite{stauffer}, usually, focus on only one 
single species and do not incorporate the inter-species interactions 
that are, however, crucially important for their extinctions.

But, in reality, evolution is a continuous process. When monitored at 
short intervals and over a not-too-long period of time, the ecological 
processes dominate the visible changes in the eco-system. However, if 
the same system is watched over sufficiently long period of time, the 
macro-evolutionary changes cannot be missed. Therefore, it is desirable 
to have a ``unified'' theoretical model that would unfold the natural 
continuous process while simulated on a computer. Very recently, 
attempts have been made by several groups to merge population dynamics 
and ``macro''-evolution within a single mathematical framework 
\cite{csk,cspre,hall,simone,per,dudek}. These efforts may have been made 
feasible, at least in part, because of the availability of fast computers.
In our models \cite{csk,cspre} we have achieved not only the merger of 
population dynamics and ``macro''-evolutionary processes but also 
detailed description of birth, ageing and death of individual organism 
so that the age-distributions in the populations of different species 
can also be monitored. 

Moreover, for the study of population dynamics of the species in the 
eco-system one needs a model of the food web, a graphic description 
of prey-predator relations \cite{cohenprs,foodweb,foodweb1,foodweb2}. 
More precisely, a food web is a directed graph where each node is 
labelled by a species' name and each directed link indicates the 
direction of flow of nutrient (i.e., {\it from} a prey {\it to} one 
of its predators). In contrast to most of the contemporary models 
published in the physics literature (ref.\cite{amaral} is one of the 
few exceptions), we incorporate the trophic level structures of real 
food webs through a generic hierarchical model \cite{cspre}. Besides, 
in order to capture ecological and evolutionary processes within the 
same theoretical framework, we allow the food web to evolve slowly with 
time \cite{thompson}.

\section{The network model: component and motivations}

\subsection{Architecture of the network} 

We model the eco-system as a dynamic {\it network} each node of which 
represents a niche that can be occupied by at most one species at a time. 
The network considered in our earliest formulation of the "unified" model 
\cite{csk} can be schematically represented by a {\it random} network. 
In a subsequent paper \cite{cspre} we replaced the random architecture 
by a generic {\it hierarchical} one, where niches are arranged in 
different trophic levels, with biologically realistic inter-species 
interactions. The hierarchical architecture helps 
us in capturing a well known fact that in the normal ecosystems the 
higher is the trophic level the fewer are the number of species. 

We assume only one single species at the highest level $\ell = 1$. There 
are $m^{\ell-1}$ levels at the ${\ell}$-th level where $m$ is a positive 
integer. The allowed range of ${\ell}$ is $1 \leq {\ell} \leq {\ell}_{max}$. 
At any arbitrary instant of time $t$ the model consists of $N(t)$ 
{\it species} each of which occupies one of the nodes of the dynamic 
network. The total number of species cannot exceed 
$N_{max} = (m^{{\ell}_{max}}-1)/(m-1)$, the total number of nodes.  
Our model allows $N(t)$ to fluctuate with time over the range 
${\ell} \leq N(t) \leq N_{max}$. The population (i.e., the total number 
of organisms) of a given species, say, $i$, at any arbitrary instant of 
time $t$ is given by $n_i(t) \leq n_{max}$. Thus, the total number of 
organisms $n(t)$ at time $t$ is given by $n(t) = \sum_{i=1}^{N(t)} n_i(t)$. 
Note that $\ell_{max}$, $m$ (and, therefore, $N_{max}$) and $n_{max}$ are 
time-independent parameters in the model. 

The network itself evolves slowly over sufficiently long time scales. 
For example, random genetic mutations are captured by implementing 
random tinkering of some of the intra-node characteristics which will 
be introduced in the next subsection. 
The inter-node interactions change slowly to capture adaptive co-evolution 
of species thereby altering the graph that represents the network.
Even the occupants of the nodes can change with time because, as the 
eco-system evolves, the populations of some species would drop to zero, 
indicating their extinction, and the corresponding nodes would be slowly 
re-occupied by new species through the process of speciation.

\subsection{Intra-node characteristics, mutations and dynamics}

The faster dynamics within each node captures ``micro''-evolution, i.e., 
the birth, growth (ageing) and natural death of the individual organisms. 
For simplicity, we assume the reproductions to be {\it asexual}.
An arbitrary species $i$ is {\it collectively} characterized by \cite{csk}:\\

\noindent (i) the {\it minimum reproduction age} $X_{rep}(i)$,\\ 
(ii) the {\it birth rate} $M(i)$,\\
(iii) the {\it maximum possible age} $X_{max}(i) = 100 \times 2^{(1-\ell)/2}$ 
that depends only on the traophic level occupied by the species. \\ 

An individual of the species $i$ can reproduce only 
after attaining the age $X_{rep}(i)$. Whenever an organism of 
this species gives birth to offsprings, $M(i)$ of 
these are born simultaneously. None of the individuals of this
species can live longer than $X_{max}(i)$, 
even if an individual manages to escape its predators. The explicit form 
of $X_{max}$ assumed above is intended to mimic the fact that the species 
at the higher trophic level usually have higher lifespan.

During each time step, because of random genetic mutations, $X_{rep}$ 
and $M$ independently increase or decrease by unity, with equal probability, 
$p_{mut}$.  $X_{rep}$ is not allowed to exceed a predetermined (large) 
positive integer while $M$ is restricted to remain positive.

The {\it intra}-species competitions among the organisms of the same 
species for limited availability of resources, other than food, imposes 
an upper limit $n_{max}$ of the allowed population of each species. 
In order to capture this requirement, we assume the {\it time-dependent} 
probability $p_b(i,\alpha)$ (of individual $\alpha$ in species $i$) 
of giving birth per unit time to be a product of two factors; one of 
the factors is a standard Verhulst factor $ 1 - n_i/n_{max}$ whereas the 
other factor takes into account the age-dependence of $p_b(i,\alpha)$. 
For an organisms, this second factor becomes non-zero only when it 
attains the corresponding minimum reproductive age $X_{rep}$; this 
non-zero factor is assumed to be of the form 
$(X_{max}-X)/(X_{max}-X_{rep})$. Note that the latter factor is unity 
at $X = X_{rep}$, and thereafter decreases, linearly with age, to zero at 
$X_{max}$. Thus  in the limit of vanishingly small population, i.e., 
$n_i \rightarrow 0$, we have $p_b(i,\alpha) \rightarrow 1$  
if $X(i,\alpha) = X_{rep}(i)$ and, thereafter, $p_b$ decreases linearly  
\cite{austad} as the organism grows older. However, 
$p_b(i,\alpha;t) \rightarrow 0$ as $n_i(t) \rightarrow n_{max}$, 
irrespective of the age of the individual organism $\alpha$ \cite{cebrat}.
Occasionally, random mutations mentioned above can lead to anomalous 
situations where some organisms may have $X_{rep} > X_{max}$; for such 
organisms $p_b(i,\alpha;t) = 0$ for all $t$ and these fail to reproduce 
during their entire life time. 

Similarly, we assume the probability $p_d$ of ``natural'' death (due to
ageing) to be a constant
$p_d = \exp[-r(X_{max}- X_{rep})/M] $, where $r$ is a small fraction, 
so long as $X < X_{rep}$. However, for $X > X_{rep}$, the probability 
of natural death increases exponentially (following Gompertz law) as 
$p_d = \exp[-r(X_{max}- X)/M] $. 
Note that, for a given $X_{max}$ and $X_{rep}$,
the larger is the $M$ the higher is the $p_d$ for any age $X$.
Therefore, in order maximize reproductive success, each species has a 
tendency to increase $M$ for giving birth to larger number of offsprings 
whereas the higher mortality for higher $M$ opposes this tendency 
\cite{tradeoff}. 

Because of the natural death mentioned above and, more importantly, 
prey-predator interactions (to be described in the next subsection),
the populations of some species may fall to zero. The nodes left 
empty by such extinct species are then re-filled by new species. 
In order to capture this process of {\it speciation}, all the empty 
nodes  in a trophic level of the network are re-filled by random 
mutants of {\it one common ancestor} which is picked up randomly from 
among the non-extinct species at the same trophic level. The subsequent 
accumulation of random mutations over sufficiently long time leads 
to the divergence of  the genomes of the parent and daughter species 
that is an essential feature of speciation.

However, occasionally, all the niches at a level may lie vacant. 
Under such circumstances, all these vacant nodes are to be filled 
by a mutant of the non-extinct species occupying the closest {\it 
lower} level. In our computer simulations, the search for this 
non-extinct species is carried out in steps, if even the lower 
level is also completely empty, the search for survivor shifts to 
the next lower level and the process continues till the lowest 
level is reached. If all the nodes, starting from the lowest, 
upto a certain level ever fall vacant, then, no new speciation 
takes place and the starvation deaths of the species propagate 
up the layers ending, finally, with the collapse of the entire 
eco-system.

\subsection{Inter-node interactions and dynamics}

The interaction between any two species $i,k$ that occupy two adjacent 
trophic levels is given by $J_{ik}$. The sign of $J_{ik}$ gives 
the direction of trophic flow, i.e. it is $+1$ if $i$ eats $k$ and 
it is $-1$ if $k$ eats $i$. In the absence of any prey-predator 
interaction between the species $i$ and $k$, $J_{ik} = 0$. For 
simplicity, we assume the absolute value (magnitudes) of all the 
non-vanishing interactions to be unity. 
Note that although there is no direct interaction between species at 
the same trophic level in our model, they can compete, albeit 
indirectly, with each other for the same food resources available in 
the form of prey at the next lower trophic level.

The $J$ account not only for the {\it inter}-species interactions but 
also {\it intra}-species competitions for food. Let $S_i^+$ be the 
number of all prey individuals for species $i$ on the lower trophic 
level, and $S_i^-$ be $m$ times the number of all predator individuals 
on the higher trophic level. Because of the larger body size of the 
predators \cite{cohen03}, we assume that a predator eats $m$ prey per 
time interval. Then, $S_i^+$ gives the available food for species $i$, 
and $S_i^-$ is the contribution of species $i$ to the available food 
for all predators on the next higher level. If the available food 
$S_i^+$  is less than the requirement, then some organisms of the 
species $i$ will die of {\it starvation}, even if none of them is 
killed by any predator. 

If $n_i-S_i^+$ is larger than $S_i^-$ then food shortage will be the 
dominant cause of premature death of a fraction of the existing 
population of the species $i$. On the other hand, if $S_i^- > n_i-S_i^+$, 
then a fraction of the existing population will be wiped out primarily 
by the predators. 

It is well known that each species tries to minimize predators but, 
at the same time, looks for new food resources. In order to capture 
this, at each time step, each of the species in our model, with the 
probability $p_{mut}$, re-adjusts a link $J$ from one of its predators and 
another to one of its potential preys \cite{sole}. If the link $J_{ij}$ 
to the species $i$ from a {\it higher} level species is non-zero, it 
is reassigned a new value $J_{ij} = J_{ji} = 0$. On the other hand, 
if the link $J_{ik}$ to a species $i$ from a {\it lower} level species 
$k$ was zero, the new values assigned are $J_{ik} = 1, J_{ki} = -1$.

\subsection{Initial conditions and update rules} 

The requirements of computational resources increase exponentially with 
increasing $\ell_{max}$. Therefore, in almost all our simulations we 
chose $\ell_{max} = 5$, although we verified that the qualitative 
features of the data were similar in case of smaller $\ell_{max}$. 
In our simulations, we always began with a random initial condition 
where $M = 1$ for all species. Since larger species occupy the higher 
tropic levels and are expected to live longer than those at lower 
levels, we assinged $X_{max} = 100, 71, 50, 35, 25$ to the species at 
level ${\ell} = 1, 2, 3, 4, 5$, respectively. Initially, $X_{rep}$ was 
assigned randomly between $1$ and $X_{max}$, the population randomly 
between $1$ and $n_{max}/2$. The ages of the individuals in the initial 
state varied randomly between $1$ and the $X_{max}$ of the corresponding 
species.

\begin{figure}[tb]
\begin{center}
\includegraphics[angle=-90,width=0.9\columnwidth]{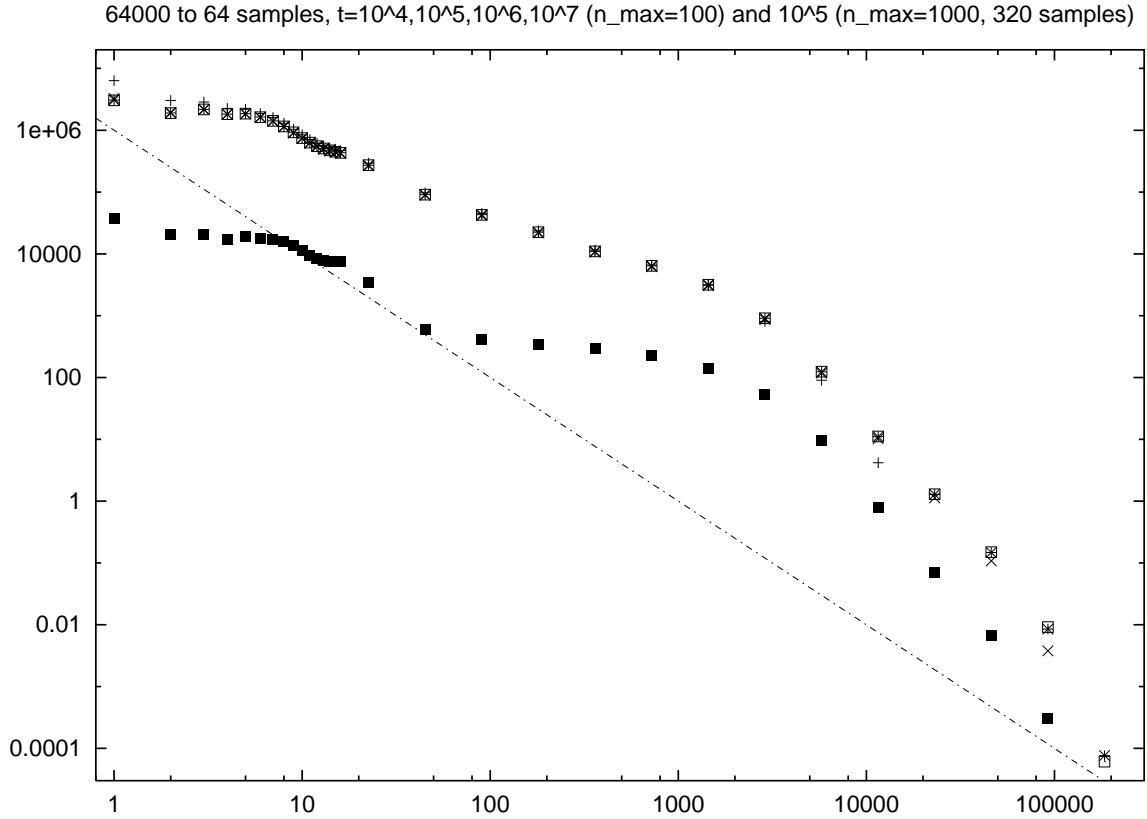}
\end{center}
\caption{Log-log plots of the distributions of the lifetimes of the 
species. The common parameters for all the curves are 
$m=2, {\ell} = 5$ (i.e. $N_{max} = 31$),  
$p_{sp} = 0.1, p_{mut} = 0.001, C = 0.05, r = 0.05$. The symbols 
$ +, \times$ and $\ast$ correspond to
$t = 10^4, 10^5, 10^6, 10^7$ averaged over $64000, 6200, 640$ and $64$ systems
respectively. The upper curves are all for $n_{max} = 100$ whereas 
the lower curve is for $n_{max} = 1000$.
The line with slope $-2$ corresponds to a power law distribution 
that has been predicted by many theories. 
}
\label{fig-1}
\end{figure}

The state of the system is updated in discrete time steps as follows: 

\noindent {\it Step I- Birth}: Each individual organism $\alpha$ 
($\alpha = 1,...,n_i(t)$) of the species $i$ ($i=1,2,...N(t)$) is 
allowed to give birth to $M(i;t)$ offsprings at every time step $t$ 
with probability (per unit time) $p_b(i,\alpha;t)$ the explicit form 
of which has been mentioned above. 

\noindent {\it Step II- Natural death}: At any arbitrary time step $t$ 
the probability (per unit time) of ``natural'' death (due to ageing) of 
an individual organism $\alpha$ of species $i$ is $p_d(i,\alpha;t)$. 

\noindent {\it Step III- Mutation}: With probability $p_{mut}$ per unit 
time, mutations of intra-node characteristics and the interactions $J$ 
are implemented.

\noindent {\it Step IV- Starvation death and killing by prey}: 

At every time step $t$, in addition to the natural death due to ageing, 
a further reduction of the population by  
\begin{equation}
C ~~\max(S_i^-,~n_i- S_i^{+})
\label{eq-kill}
\end{equation}
is implemented where $n_i(t)$ is the population of the species $i$ 
that survives after the natural death step above. $C$ is a constant 
of proportionality. If implementation of these steps makes $n_i \leq 0$, 
species $i$ becomes extinct.  

\noindent {\it Step V- Speciation}: The nodes left empty by extinction 
are re-filled by new species, with probability $p_{sp}$ following 
the algorithm for speciation mentioned above. 

The longest runs in our computer simulations were continued upto  
$10^8$  time steps. If each time step in our model is assumed to 
correspond to a real time of the order of one year, then the total 
time for which we have monitored our model eco-system, is comparable 
to real geological times scales.

\begin{figure}[tb]
\begin{center}
\includegraphics[angle=-90,width=0.9\columnwidth]{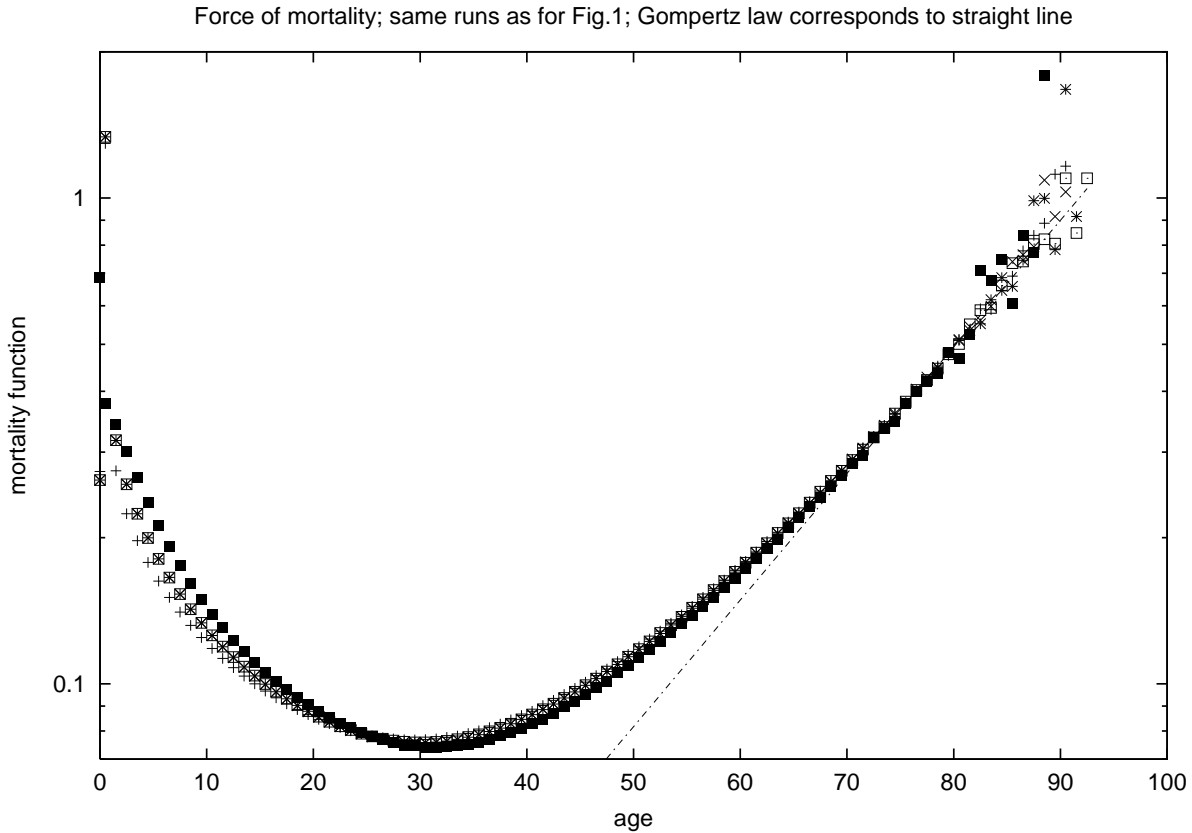}
\end{center}
\caption{Semi-log plot of the mortality function against age of the 
individual organisms at the highest trophic level. The same symbols 
in fig.\ref{fig-1} and fig.\ref{fig-2} correspond to the same set 
of parameter values.
}
\label{fig-2}
\end{figure}

\section{Results} 

\subsection{Lifetime distributions}

The average distributions of the lifetimes of the species are plotted 
in fig.\ref{fig-1}. It is not possible to fit a straight line through 
the data over the entire range of lifetimes although only a limited 
regime is consistent with a power-law with the effective exponent 
$2$, which has been predicted by several models of ``macro''-evolution 
\cite{drossel,newman}. The overlap of the curves for different simulation 
times establishes that our simulations have reached 
the asymptotic regime where effects of initial conditions have been 
completely washed out. 
Because of the various known limitations of the available fossil data, 
it is questionable whether real extinctions follow power laws and, if 
so, over how many orders of magnitude. 

\begin{figure}[tb]
\begin{center}
\includegraphics[angle=-90,width=0.9\columnwidth]{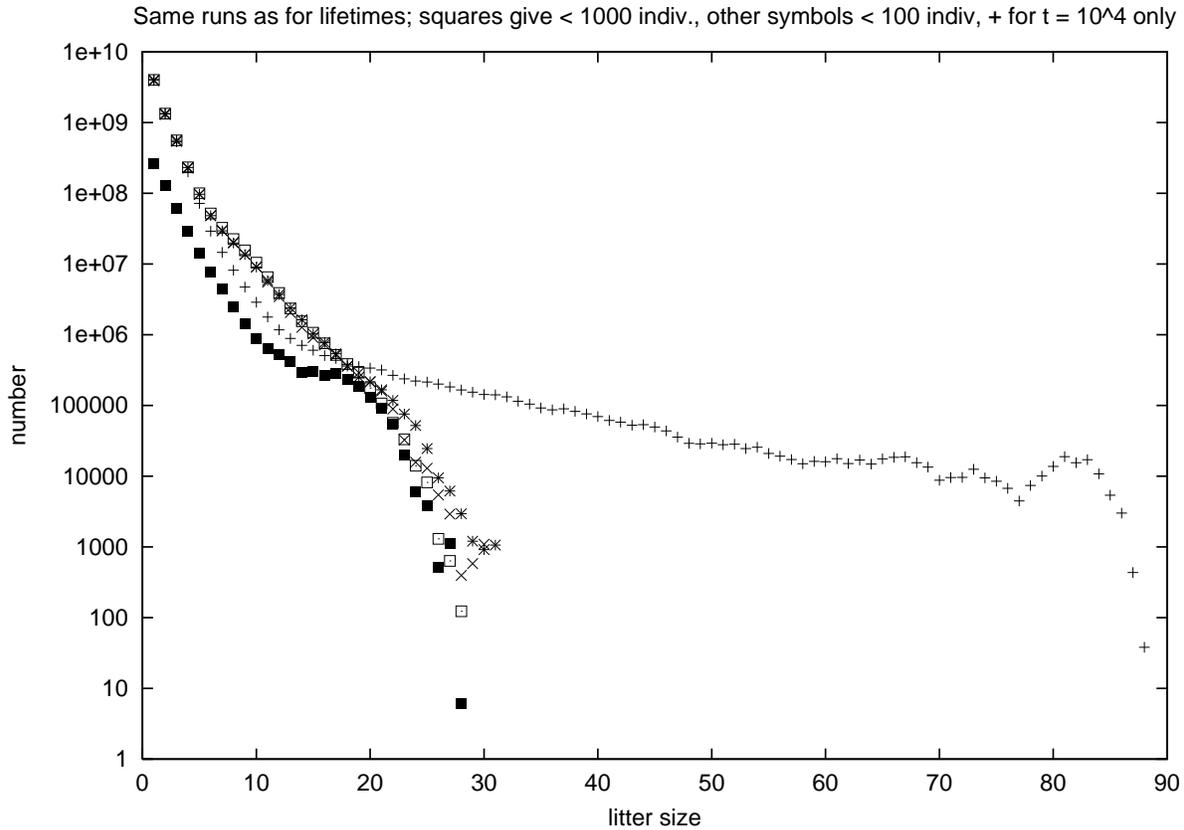}
\end{center}
\caption{Semi-log plot of the distribution of $M$. The same symbols 
in fig.\ref{fig-1} and fig.\ref{fig-3} correspond to the same set 
of parameter values.
}
\label{fig-3}
\end{figure}

\subsection{Distributions of Species Characteristics}

We define the mortality rate by the relation $-d~~\ln(survivors)/d(age)$. 
The mortality rate extracted from the raw data is plotted against age 
in fig.\ref{fig-2}. The shape of this curve is consistent with the 
usual census data that show a minimum in childhood and exponential 
increase in adults.

During the early stages of the macro-evolution, the distributions of $M$ 
broadens. But, with further passage of time it shrinks and reaches 
a stationary form where the largest $M$ is around $30$ 
(for the parameter set used in our simulations). This 
is consistent with the view that organisms have a choice of either 
faster reproduction and shorter life  or slower reproduction and longer 
life span.

\subsection{Collapse of fragile ecosystems}

One crucial effect of the generalization of the rule for speciation 
proposed in this paper is that the eco-systems are now much more 
stabilized than in our earlier papers \cite{csk,cspre}. In our 
earlier work, reported in ref.\cite{cspre} we allowed an empty node 
to be re-occupied by one of the non-extinct species from only the 
same level where the extinct node was located. If, by chance, all 
the nodes of one level fell vacant at some stage, no speciation could 
take place; this would trigger an avalanche of extinctions that would 
propagate upward in the food chain because of starvation of the organisms 
and eventually lead to a collapse of the entire eco-system. In contrast, 
in the current version of our model, speciation is allowed to take 
place from lower levels if all the nodes in an entire level become 
vacant. This reduces the possibility of collapse of the eco-system 
but does not rule it out completely. Work on further stabilization of 
the eco-system with variable ${\ell}_{max}$ is in progress and will be 
reported elsewhere in a future publication \cite{csbio}.

\section{Summary and conclusion}

In summary, we have extended our dynamic network model for unified 
description of micro- and macro-evolution. A majority of the main 
characteristics of the model are emergent properties of the 
self-organizing dynamics of the system. The main effect of the 
extension of the speciation dynamics proposed in this paper is 
that even after large avalanches of extinctions the eco-system can 
recover by speciation and bio-diversification starting from the 
surviving species at the low trophic levels. Consequently, the 
complete collapse of the eco-system becomes extremely rare and 
the distribution of the litter size $M$ became stationary much 
faster.

\noindent {\bf Acknowledgements} 

We thank J.E. Cohen for useful correspondence and the Supercomputer 
Center J\"ulich for computer time on their CRAY-T3E. This work is 
supported by Deutsche Forschungsgemeinschaft (DFG) through a 
Indo-German joint research project. 

\bibliographystyle{plain}

\begin{thebibliography}{1}

\bibitem{hopfield} L.H. Hartwell, J.J. Hopfield, S. Leibler and A.W. Murray, 
Nature {\bf 402}(Suppl.), C47 (1999). 

\bibitem{bornholdt} S. Bornholdt and H. G. Schuster (eds.), 
{\it Handbook of Graphs and Networks - From the Genome to the 
Internet}, (Wiley-VCH, Weinheim, 2003).

\bibitem{foodweb2} Drossel B and McKane, A.J., Modelling food webs, p. 
218 in ref.\cite{bornholdt}. 

\bibitem{drossel} B. Drossel,  Adv. Phys. {\bf 50}, 209 (2001). 

\bibitem{newman} M. E. J. Newman and R. G. Palmer,  adap-org/9908002;
{\it Modeling Extinction}, (Oxford University Press, New York, 2002);

\bibitem{lassig} M. L\"assig and A. Vallerian, (eds.) {\it Biological Evolution and Statistical Physics}, (Springer Verlag, Berlin-Heidelberg 2002)

\bibitem{csk} D. Chowdhury, D. Stauffer and A. Kunwar, Phys. Rev. Lett. 
{\bf 90}, 068101 (2003).

\bibitem{cspre} D. Chowdhury and D. Stauffer, cond-mat/0305322.

\bibitem{murray} J.D. Murray, {\it Mathematical Biology}, (Springer, 1989)

\bibitem{lotka} N.S. Goel, S.C. Maitra and E.W. Montroll, Rev. Mod. Phys. {\bf 43}, 231 (1971); J. Hofbauer and K. Sigmund, {\it The Theory of Evolution and Dynamical Systems} (Cambridge University Press, 1988).

\bibitem{kauffman} S. Kauffman, {\it The Origins of Order: Self-organization 
and selection in Evolution} (Oxford University Press, New York, 1993).

\bibitem{bak} P. Bak and K. Sneppen, Phys. Rev. Lett. {\bf 71}, 4083 (1993); 
M. Paczuski, S. Maslov and P. Bak, Phys. Rev. E {\bf 53}, 414 (1996). 

\bibitem{solerev} R. V. Sole, in: {\it Statistical Mechanics of Biocomplexity}, eds. D. Reguera, J. M. G. Vilar and J. M. Rubi, Lec. Notes in Phys. vol.527, 217 (Springer, 1999)
 
\bibitem{stauffer} D. Stauffer, in ref.\cite{lassig}. 

\bibitem{hall} M. Hall, K. Christensen, S. A. di Collobiano and H.J. Jensen, Phys. Rev. E {\bf 66}, 011904 (2002) 

\bibitem{simone} S. A. di Collobiano, K. Christensen and H.J. Jensen, J. Phys. A {\bf 36}, 883 (2003). 

\bibitem{per} P. A. Rikvold and R. K. P. Zia,  nlin.AO/0303010, nlin.AO/0306023.

\bibitem{dudek} A. Nowicka, A. Duda and M. R. Dudek,  cond-mat/0207198;
A. {\L}aszkiewicz, Sz. Szymczak, S. Cebrat, Int. J. Mod. Phys. C 14, issue 6
(2003)

\bibitem{cohenprs} J. E. Cohen, T. Luczak, C. M. Newman and Z. -M Zhou, Proc. Roy. Soc. Lond. B {\bf 240}, 607 (1990).

\bibitem{foodweb} G. A. Polis and K. O. Winemiller (eds.)  {\it Food 
Webs: Integration of Patterns and Dynamics} (Chapman and Hall, New York, 1996). 

\bibitem{foodweb1} S. L. Pimm, {\it Food Webs} (Chapman and Hall, London, 1982).

\bibitem{amaral} L. A. N. Amaral and M. Meyer, Phys. Rev. Lett. {\bf 82}, 652 (1999). 

\bibitem{thompson} J. N. Thompson, Science {\bf 284}, 2116 (1999).

\bibitem{austad} S.N. Austad, in: {\it Between Zeus and the Salmon: The 
Biodemography of Longevity}, K. W. Wachter, and C. E. Finch, eds.  
(National Academy Press, Washington DC, 1997). 

\bibitem{cebrat} J. S. S\'a Martins, and S. Cebrat, Theory in Biosciences 
{\bf 199}, 156 (2000).

\bibitem{tradeoff} C. K. Ghalambor,  and T. E. Martin, Science {\bf 292}, 
494 (2001).

\bibitem{cohen03} J. E. Cohen, T. Jonsson and S. R. Carpenter, Proc. 
Natl. Acad. Sci. USA {\bf 100}, 1781 (2003). 

\bibitem{sole} R. V. Sol\'e  and S. C. Manrubia, Phys. Rev. E 
{\bf 54}, R42 (1996); {\bf 55}, 4500 (1997).

\bibitem{csbio} D. Chowdhury and D. Stauffer, (to be published).



\end{thebibliography}

\end{document}